\documentclass[RNAAS]{aastex63}

\graphicspath{{./}{figures/}}
\usepackage[caption=false]{subfig}
\usepackage{amsmath}
\usepackage{hyperref}

\begin{document}

\title{Fisher Matrix Stability}


\author{Naren Bhandari}
\altaffiliation{nbhandar@andrew.cmu.edu}
\affiliation{McWilliams Center for Cosmology, Department of Physics, Carnegie Mellon University, \\
Pittsburgh, PA 15213, USA}

\author{C. Danielle Leonard}
\altaffiliation{danielle.leonard@ncl.ac.uk}
\affiliation{School of Mathematics, Statistics and Physics, Newcastle University, Newcastle upon Tyne, NE1 7RU, UK}

\author{Markus Michael Rau}
\altaffiliation{markusmichael.rau@gmail.com}
\affiliation{McWilliams Center for Cosmology, Department of Physics, Carnegie Mellon University, \\
Pittsburgh, PA 15213, USA}

\author{Rachel Mandelbaum}
\altaffiliation{rmandelb@andrew.cmu.edu}
\affiliation{McWilliams Center for Cosmology, Department of Physics, Carnegie Mellon University, \\
Pittsburgh, PA 15213, USA}


\section{Abstract}
\label{sec:abstract}
Fisher forecasts are a common tool in cosmology with applications ranging from survey planning to the development of new cosmological probes. While frequently adopted, they are subject to numerical instabilities that need to be carefully investigated to ensure accurate and reproducible results. This research note discusses these challenges using the example of a weak lensing data vector and proposes procedures that can help in their solution.

\section{Introduction} \label{sec:intro}
With the advent of ongoing large photometric surveys like the Dark Energy Survey \citep[DES;][]{DES}, the Kilo-Degree Survey \citep[KiDS;][]{2013ExA....35...25D}, the Hyper Suprime-Cam survey \citep[HSC;][]{2018PASJ...70S...4A} and upcoming surveys like the Vera C.\ Rubin Observatory Legacy Survey of Space and Time \citep[LSST;][]{lsst_desc_outcomes} and {\it Euclid} \citep{Euclid}, the need for computationally efficient forecasting for survey planning or to investigate the impact of sources of systematic error increases. Fisher forecasts are a commonly used tool for this purpose, that has been adopted early in many areas of astrophysics and cosmology \citep[e.g.][]{BASSETT_2011,1994PhRvD..49.2658C, 1996PhRvD..54.1332J, 1996PhRvL..76.1007J, Tegmark_1997_surveys, Tegmark_1997_fisher, 2020MNRAS.495.4210H} to forecast parameter constraints without the barrier of the computational expense of e.g. Markov chain Monte Carlo methods. However, in making use of them, there are issues of which we must be aware. We highlight the importance of rigorous investigations of numerical stability in Fisher forecasts while using a weak lensing convergence power spectrum data vector. We discuss possible pitfalls in the application of Fisher forecasts and propose methodological tests to ensure numerical accuracy. Our work builds on that of \cite{Yahia2020}, where this question was considered for the complementary case of a data vector consisting of the galaxy power spectrum measured from a spectroscopic galaxy survey (with a specific focus on {\it Euclid}), and on that of \cite{Euclidforecasts}, which performed an analytic estimation of the relationship between the maximum fractional error on an element of the Fisher matrix and the error on the elements of the resulting forecast parameter covariance matrix.

In the course of this investigation we will use two libraries that provide the capability to model weak lensing convergence power spectra: CCL v2.0.1 \citep{CCL_2019} and CosmoSIS v1.6 \citep{Zuntz_2015}. We note that the settings for CosmoSIS that we use in this work are optimized for MCMC sampling and not for high accuracy Fisher matrix calculations. This is not a limitation of the code itself, since higher accuracy configurations are supported. The purpose of this work is not to perform a code comparison, but rather to investigate the importance and impact of numerical derivative calculations on the reliability of Fisher forecasts, and to provide tests that can be used to avoid numerical pitfalls. The CosmoSIS code package has been subject to significant scrutiny in terms of the accuracy of likelihood evaluations in e.g.~\citet{2017arXiv170609359K}.  

The structure of this research note is as follows.  In the `\nameref{sec:proc}' section, we outline our fiducial assumptions and how we set up our Fisher forecast. In the `\nameref{sec:results}' section, we detail our findings using both libraries and gauge their numerical stability in the context of Fisher forecasting. Finally, in the `\nameref{sec:conc}', we summarize and present our conclusions regarding these forecasts.

\newpage
\section{Forecast Methodology} \label{sec:proc}
\begin{figure}
    \centering
    \includegraphics[width=0.55\columnwidth]{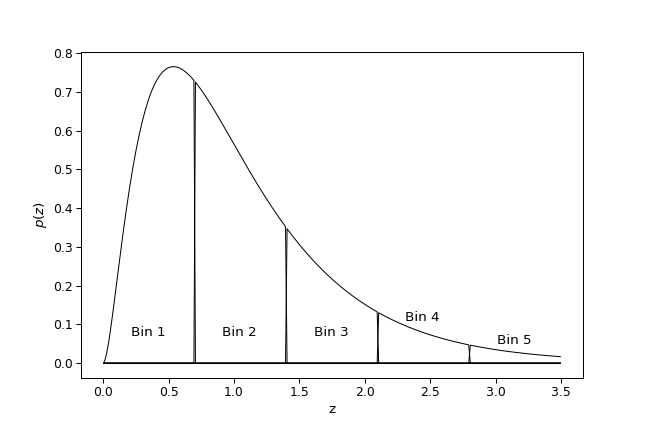}
    \caption{Redshift probability density function, $p(z)$, as a function of redshift, $z$.}
    \label{fig:pz_bin}
\end{figure}

We perform a Fisher forecast for an LSST-like convergence power spectrum data vector assuming the fiducial cosmology and survey setup summarized in Table~\ref{tab:cosmo_setup}. We use $\theta = (\Omega_c, w_0, h, 10^9 A_s, \Omega_b, n_s, w_a)$ as our cosmological parameter vector and we use a redshift distribution\footnote{We use the \texttt{zdistri\_model\_z0=1.100000e-01\_beta=6.800000e-01\_Y10\_source} redshift distribution} from the LSST Dark Energy Science Collaboration (DESC) Science Requirements Document v1 \citep{lsst_srd_zenodo, lsst_srd} which we bin into five tophat tomographic bins that are equally spaced in redshift. Figure \ref{fig:pz_bin} shows the redshift probability density function, $p(z)$, i.e. the probability that a randomly selected galaxy has redshift $z$, along with the five tophat selections. Furthermore, we take the fractional sky coverage, $f_{\rm{sky}}$, from Table~1.1 in \cite{lsst_science_book}, and choose the number density, $\overline{n}_g$, to be consistent with Table~1 in \cite{eff_num_dens}.

\begin{deluxetable}{lR|R}[ht]
\tablecaption{Fiducial Parameter Values and Survey Assumptions\label{tab:cosmo_setup}}
\tablehead{
\colhead{Cosmological/Survey Parameter}  & & \colhead{Value}}
\startdata
Dark Matter Density                                 & \Omega_c  & 0.25\\
Baryon Density                                      & \Omega_b  & 0.04\\
Dark Energy Equation of State                       & w_0       &-0.90\\
\qquad\qquad $w(a)=w_0+w_a(1-a)$                    & w_a       & 0.00\\
Reduced Hubble Constant                             & h       & 0.72\\
Primordial Power Spectrum Amplitude \quad\quad    & 10^9 A_s  & 2.10\\
Slope of Primordial Power Spectrum                  & n_s       & 0.96\\
Shape Noise                                         & \sigma_{\epsilon}^2 & 0.23\\
Fractional Sky Coverage                             & f_{\rm{sky}} & 0.48\\
Number of Galaxies per Steradian                    & \overline{n}_g & 3.1\times 10^8
\enddata
\end{deluxetable}

\paragraph{Convergence Power Spectrum Modelling}
Using CCL and CosmoSIS, we calculate the shear-shear angular power spectra which serve as our data vector.
The convergence power spectrum $ C^{\kappa}(\ell)$ probes the spatial correlation of the lensing convergence signal \citep[e.g.][]{1991MNRAS.251..600B, 1992ApJ...388..272K, 2001PhR...340..291B}
\begin{equation}
 C^{\kappa}(\ell) = \frac{9}{4} \Omega_{m}^2 H_0^4 \int_{0}^{\chi_H} \mathrm{d}\, \chi \,  \frac{q^2(\chi)}{a^2(\chi)} P_{\rm \delta}\left(k = \frac{\ell}{f_K(\chi)}, \chi\right) \, .
 \label{eq:convergence_power_spectrum}
\end{equation}
Here, $\Omega_m$, $H_0$ denote the matter density and Hubble constant respectively, $\chi_H$ is the comoving distance $\chi$ for which $\chi(a) \rightarrow \chi_H$ as the scale factor $a \rightarrow 0$. Note that $\Omega_m = \Omega_b + \Omega_c$.
From the structure of Eq.~\ref{eq:convergence_power_spectrum} we see that the matter power spectrum $P_{\rm \delta}\left(k = \frac{\ell}{f_K(\chi)}, \chi\right)$ is projected along the line-of-sight using a geometrical factor, the `lensing weight': 
\begin{equation}
 q(\chi) = \int_{\chi}^{\chi_H} \mathrm{d}\, \chi^{'} \, n(\chi^{'}) \left(\frac{f_K(\chi^{'} - \chi)}{f_K(\chi^{'})}\right) \, ,
 \label{eq:lens_weight}
\end{equation}
where the function $f_K(\chi)$ is a curvature-dependent factor in the FLRW metric. We see that this geometrical factor critically depends on the redshift distribution $n(\chi^{'})$ that constitutes an important source of systematic uncertainty in weak lensing measurements. For a more detailed treatment of the convergence power spectrum we refer the reader to the aforementioned references. For the matter power spectrum $P_\delta$, we use the \citet{Takahashi_2012} version of the nonlinear power spectrum from Halofit \citep{Smith_2003}.

\paragraph{Covariance Matrix Modelling}
As in \cite{cov_donnacha_kirk}, we use the convergence power spectrum to calculate the covariance matrix. The Gaussian model for the covariance matrix of the data vector is given as
\begin{equation}
    \Sigma^{(k, m)}_{(i, j)} (\ell) = \frac{1}{(2 \ell + 1)f_{\rm sky}} \left( \overline{C}^{(i, k)}(\ell) \overline{C}^{(j, m)}(\ell) + \overline{C}^{(i, m)}(\ell) \overline{C}^{(j, k)}(\ell) \right) \, ,
\end{equation}

where
\begin{equation}
    \overline{C}^{(i, j)}(\ell) = C^{(i, j)}(\ell) + \delta_{i, j} \, \left(\frac{\sigma_{\epsilon}^2}{2 \, \overline{n}_g^{i}}\right) \, .
\end{equation}
Here, $\overline{n}_{g}^{i}$ denotes the number of galaxies per steradian and $\sigma_{\epsilon}$ is the shape noise expressed as a standard deviation.

Note that the indices denote samples selected in bins of increasing redshift, i.e. tomographic bins, whose redshift probability density functions are shown in Fig.~\ref{fig:pz_bin}. The index combinations $(i, j)$ denote the respective auto- ($i = j$) and cross-correlations ($i \neq j$) obtained by correlating the samples in the tomographic selections $i$ and $j$. For example, at fixed $\ell$, the correlation between the auto-correlation convergence power spectrum of the first bin with the cross-correlation convergence power spectrum of first and second tomographic redshift bin, would be $\Sigma^{(k=1, m=1)}_{(i = 1, j = 2)}$. At each $\ell$, the data-vector, denoted by $C(\ell)$, is constructed from all convergence power spectra formed from the available tomographic redshift bins, i.e., by enumerating the possible combinations we can map the pair of indices $(i, j)$ to a single index and make a vector. Similarly, at each $\ell$, the data-covariance matrix, denoted by $\Sigma(\ell)$, is constructed such that the rows and columns are indexed by the enumeration above.

\paragraph{Fisher Forecast}
Suppose that the negative log-likelihood of our data vector is $\mathcal{L}(C(\ell);\theta)$. Then the Fisher matrix is defined as the expectation value of the Hessian matrix of $\mathcal{L}$ with respect to the parameters $\theta$:
\begin{equation} \label{eq:fisher_hess}
    F_{\alpha \beta} = \left\langle \frac{\partial ^2 \mathcal{L}}{\partial \theta_\alpha\partial \theta_\beta} \right\rangle
\end{equation}
Assuming a Gaussian likelihood and that $\Sigma(\ell)$ is independent of $\theta$, we can calculate the Fisher matrix as
\begin{equation} \label{eq:fisher}
    F_{\alpha \beta} = \sum_{\ell}\frac{\partial C(\ell)^T}{\partial \theta_\alpha} \cdot \Sigma(\ell)^{-1} \cdot \frac{\partial C(\ell)}{\partial \theta_\beta}
\end{equation}
where the derivatives are evaluated at the fiducial parameter values \citep{cov_donnacha_kirk}. The sum is taken over $\ell_{\min{}} = 76$ to $\ell_{\max{}} = 999$ --- this range is motivated by accessible values to the aforementioned surveys and also to stay in the linear and quasi-linear regime.
 We refer the reader to \cite{Tegmark_1997_surveys, Tegmark_1997_fisher, BASSETT_2011, Heavens_2014} for more detailed treatments of the Fisher matrix formalism.

\paragraph{Numerical Differentiation}
We calculate the derivatives above numerically by using centered finite difference methods outlined in \cite{fornberg1988generation}:
\begin{equation}
    \frac{\partial f}{\partial x} \bigg |_{x=x_0} = \sum_{j = - n}^n \mu_j f(x + j \Delta s)
\end{equation}
where $f$ is some function that depends on the variable $x$, $\mu_j$'s are coefficients given by the aforementioned reference, $\Delta s$ is the step-size, and $n$ is an integer that determines the number of points surrounding the reference point where the derivative needs to be approximated. The geometrical arrangement of these points is referred to as the `stencil'. The number of points in the stencil, namely $2n + 1$ for centered finite difference methods, determines how the numerical error of the stencil method scales with the step-size. For example, the five-point stencil approximation is given by 
\begin{equation}
    \frac{\partial f}{\partial x} \bigg |_{x=x_0} = 
    \frac{1}{12 \Delta s} f(x_0 - 2\Delta s) 
    -\frac{8}{12 \Delta s} f(x_0 - \Delta s)
    + 0 f(x_0)
    +\frac{8}{12 \Delta s} f(x_0 + \Delta s) 
    -\frac{1}{12 \Delta s} f(x_0 + 2\Delta s).
\end{equation}
Note that all the centered finite differences we use have a zero coefficient on the central term. As such, while entering into the theoretical modelling of the covariance matrix, the central term does not enter the calculation of the numerical derivative. Thus, the order of accuracy is given by $2n$, i.e., evaluating more points in our stencil will lead to more accurate derivatives at the cost of computational complexity (here, more calls to CCL or CosmoSIS). We will investigate the numerical performance as a function of the number of points in the stencil and the step-size. We do so by calculating derivatives using three-, five-, seven-, and nine-point stencils and seeing how some figure of merit changes in response to varying the step-size $\Delta s$, looking for a region where it is stable --- henceforth, we will refer to this process as tuning the derivatives. In this context it is noteworthy that other derivative tuning methods have been employed by the community. For example \citet{2017MNRAS.464.4747C}, due the large number of observables and cosmological parameters in their work, use a linear interpolation constructed on a set of sampling points around the fiducial value for derivative estimation. They then decrease the point spacing adaptively until the required numerical accuracy is met. In this work we will restrict our approach to a visual approach to gain understanding of the behaviour of different codes as a function of step size. For the sake of comparisons, it will be convenient to work with a normalized step-size given by the nominal step-size divided by the absolute value of the fiducial parameter value instead. For the parameter, $w_a$, which has a fiducial value of 0, we divide the nominal step size by the absolute value of the fiducial value of $w_0$, which is nonzero.

\paragraph{Performance Evaluation Metrics}
In order to gauge how well the measurement we are forecasting constrains dark energy, we use the Dark Energy Task Force's Figure of Merit (DETF FoM) given by 
\begin{equation}\label{DETF FoM}
    \text{DETF FoM} = \sqrt{\det{F^{w_0 w_a}}}
\end{equation}
where $F^{w_0 w_a}$ is the reduced Fisher matrix, obtained by inverting the Fisher matrix $F$, marginalizing over parameters that are not $w_0$ or $w_a$, and inverting back \citep{albrecht2006report}.
Note that we can also define similar quantities for other pairs of parameters, like $\Omega_c$ and $10^9 A_s$ for example. To distinguish which pairs of parameters we are talking about, we refer explicitly to the $w_0 - w_a$ FoM (as in Eq.~\ref{DETF FoM}) or $\Omega_c - 10^9 A_s$ FoM where necessary. Since the main purpose of this note is to study how numerical errors can bias Fisher matrix inference, we are not marginalizing over intrinsic alignments or other sources of systematic uncertainty.

\section{Results} \label{sec:results}

Both CCL and CosmoSIS are well-trusted and mature libraries that we would expect to produce theoretical predictions and forecasts that are consistent, so we begin by checking this assumption.
As a first test of consistency between the data vectors produced by CCL and CosmoSIS we investigate the signal-to-noise ratio (${\rm SNR}$) given by:
    \begin{equation}
     {\rm SNR} = \sqrt{\sum_{\ell} C(\ell) \cdot \Sigma(\ell)^{-1} \cdot C(\ell)}
    \end{equation}
These were calculated to be $342.58$ and $342.44$ for CCL and CosmoSIS, respectively --- a relative difference of $-0.04\% $, which we consider to be a good agreement. This first test is important to ensure that the data vector and the covariance calculation give reasonable results, before numerical differentiation introduces numerical sources of error that need to be isolated. 

\begin{figure}[h!]
\begin{center}
     \centering
     \subfloat{
      \includegraphics[width=0.48\columnwidth]{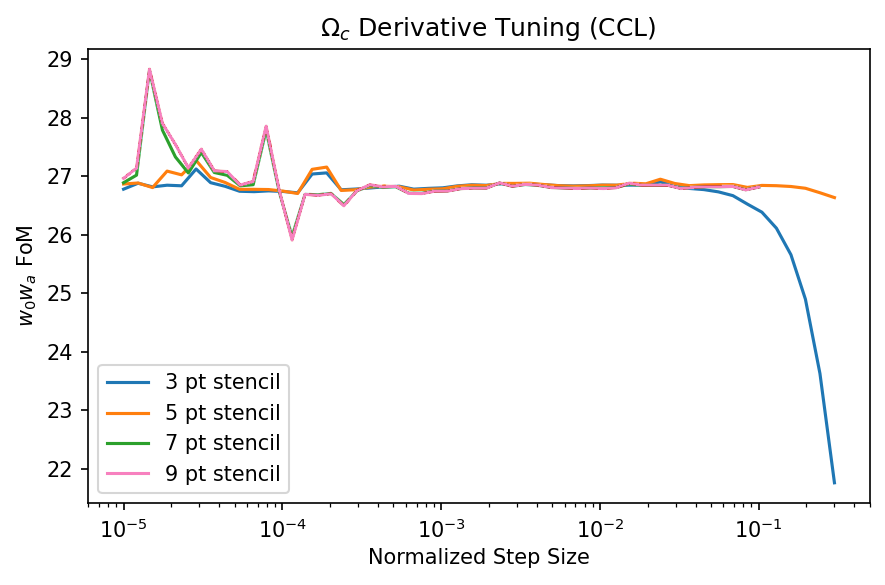}
      }
      ~
      \subfloat{
      \includegraphics[width=0.48\columnwidth]{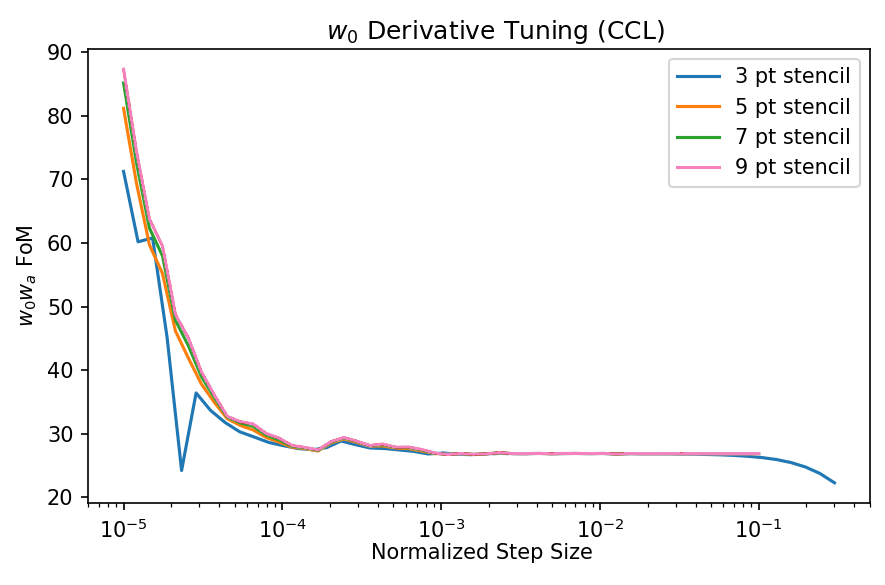}
      }

\caption{Derivative tunings for CCL, focusing on parameters $\Omega_c$ (left) and $w_0$ (right), using various order stencils. Note that a normalized step-size of 0.01 appears stable.\label{fig:ccl_tuning}}
\end{center}
\end{figure}

\begin{figure}[h!]
\begin{center}
     \centering
     \subfloat{
      \includegraphics[width=0.48\columnwidth]{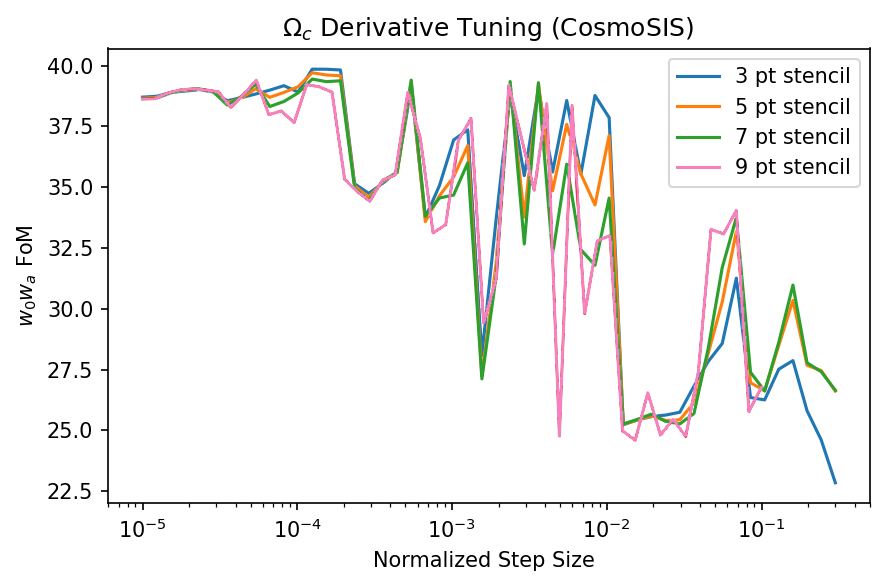}
      }
      ~
      \subfloat{
      \includegraphics[width=0.48\columnwidth]{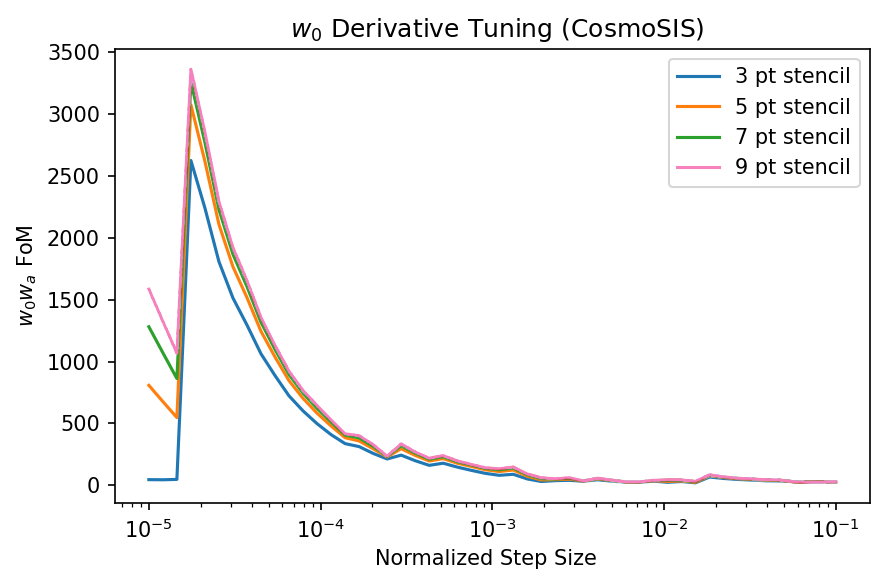}
      }
\caption{Derivative tunings for CosmoSIS, focusing on parameters $\Omega_c$ (left) and $w_0$ (right) using various order stencils. Note that there does not seem to be a clear region of stability in either case. In the latter case, this is true even restricting the range of the plot to focus on the larger normalized step-sizes. \label{fig:cosmosis_tuning}}
\end{center}
\end{figure}

In order to calculate the Fisher Matrices, we tune the derivatives as mentioned in the `\nameref{sec:proc}' Section. We select 0.01 as an initial guess for a starting normalized step-size for all parameters and calculate all the derivatives using a five-point stencil and arrive at an initial DETF FoM estimate.  
To then tune the derivatives, we calculate them using the three-, five-, seven-, and nine-point stencils,
varying the normalized step-size to look for regions where the DETF FoM does not vary much. In the CCL case, we were able to find clear regions of stability, wherein the normalized step-sizes for all parameters was chosen to be 0.01 using a nine-point stencil --- a few examples can be seen in Figure \ref{fig:ccl_tuning}. This spacing is of the same order of magnitude as that determined to be optimal for each cosmological parameter in \cite{Yahia2020} for the spectroscopic galaxy case; this makes sense as both observables rely on underlying calculations of the power spectrum via Boltzmann codes. In the CosmoSIS case, we were unable find such a region of stability for the parameters $\Omega_c$ and $w_0$, seen in Figure \ref{fig:cosmosis_tuning}. However, for the sake of further comparison, we can pick values of the normalized step-sizes and the number of points in the stencil used (like 0.02 for the nine-point stencil for $\Omega_c$) that bring the DETF FoM as close as possible to its CCL counterpart, as seen in the third column of Table~\ref{tab:summary}. (Despite this, we still see, in the next column, that the analogous FoM for $\Omega_c$ and $10^9 A_s$ are quite different between CCL and CosmoSIS). While we were unable to successfully tune derivatives for CosmoSIS, there are other methods of forecasting with the library. We note that our CosmoSIS setup is mainly optimized for MCMC techniques that may not require a high numerical precision in the calculation of angular power spectra.
Both the internal Boltzmann solvers and the CosmoSIS modules that perform the integration over the window functions can be adjusted towards increased numerical accuracy and are subject to future accuracy improvements. We  reiterate that we aren't trying to compare the numerical accuracy of the two libraries, but use them to investigate numerical issues that can arise in Fisher forecasts. 

Continuing with the calculation of our Fisher forecast, we get Fisher matrices from CCL and CosmoSIS that have entry-wise relative differences that are no larger than $0.6\%$ in magnitude as seen in Table~\ref{tab:fisher_rel_diff}. However, these Fisher matrices have high condition numbers (see Table~\ref{tab:summary}) that lead to large magnifications in the entry-wise relative differences when we invert the matrices and look at the corresponding covariance matrices as seen in Table~ \ref{tab:cov_rel_diff}. At worst, the relative difference is almost a factor of 15. This highlights how small instabilities in the Fisher matrix can translate to much larger deviations in the parameter covariance matrix. Note that we use the inversion method, \verb+linalg.inv+, implemented in the NumPy package \citep{numpy_oliphant2006, numpy_van2011} that is based on LU decomposition. In the following paragraph we will describe the consistency tests we performed to ensure numerical accuracy in the inversion. 

Given the high condition numbers, we gauge how accurate our calculated inverses are by one of two methods: 

\begin{enumerate}
    \item We multiply a Fisher matrix with its corresponding covariance matrix (its inverse) and check whether we get something that is numerically close to the identity matrix. 
    
    \item We follow \cite{kids}, which uses a technique from \cite{newman1974determine} to verify matrix inversions in its forecasts. The latter gives upper and lower bounds on the matrix norm of the difference between the calculated and actual inverse of a matrix. Let $C$ be our calculated inverse to the Fisher matrix $F$, let $F^{-1}$ be the actual inverse of $F$, let $I$ be the Identity matrix, let $R = I - FC$ and let $N$ be a matrix norm (we use the Frobenius norm). Then 
    \begin{equation}
        \frac{N(CR)}{1+N(R)} \leq N(F^{-1} - C) \leq \frac{N(CR)}{1-N(R)}
    \end{equation}
    For CCL, the lower bound was calculated to be $1.88 \times 10^{-12}$ and the upper bound differed by $1.70 \times 10^{-22}$. For CosmoSIS, the respective lower bound was $3.09 \times 10^{-13}$ and the upper bound differed by $2.28 \times 10^{-23}$.
    By subtracting the lower bound from the upper bound, we get the size of the interval that contains the norm in question. This `Inverse Norm Spread' is shown again in the last column of Table~\ref{tab:summary}. Since these spreads are very small in all cases, we have another test suggesting that our calculated inverses are accurate.
\end{enumerate}

We also test the sensitivity of our results with respect to a reparametrization wherein we use $\Omega_c h^2$ and $\Omega_b h^2$ instead of $\Omega_c$ and $\Omega_b$ respectively, and repeat the above steps. The last row of Table~\ref{tab:summary} summarize the condition number, the DETF and $\Omega_c h^2 - A_s$ FoM, and the Inverse Norm Spread for this case using CCL. While the condition number is still high, the DETF FoM is roughly the same as earlier and the small Inverse Norm Spread suggests our calculated covariance matrix is numerically accurate.

In the field of numerical stability, it is well known that when inverting a matrix, it should be equilibrated, i.e., given a norm, its rows (and columns) should have similar lengths in that norm. This ensures that small perturbations in all rows (or columns) are of similar magnitude. For a more detailed treatment of the subject, we refer the reader to \cite{stab_mat_inv, wilkinson_mat_inv, equilibration_sym}. In the context of Fisher forecasting, if one is using $A_s$ as a parameter, equilibration can be used to lower the condition number. For example, we can scale $A_s$ by $10^9$ and use $10^9 A_s$ as a parameter. While this does not affect the relative differences in the Fisher and covariance matrix, we found that it does make eigenvalue/eigenvector calculations more accurate. Since doing so comes at no computational cost, we incorporated this into our fiducial ansatz. Note that $\log [10^{10} A_s]$ has been used as well \citep{Planck_2015_data_overview, raveri2016information, Ballardini_2018}.

Additionally, we also test how adding priors on $\Omega_b$ and $h$ affects our results.
Using Monte Carlo Markov Chains provided in the 2015 release of data products\footnote{We use the chains in the \texttt{base\_plikHM\_TT\_lowTEB\_post\_BAO.txt} files from the `Planck baseline model: TT low-ell temperature and LFI polarization (bflike, 2 \textless= ell \textless= 29)' data available at \url{https://pla.esac.esa.int/pla/aio/product-action?COSMOLOGY.FILE_ID=COM_CosmoParams_base-plikHM-TT-lowTEB_R2.00.tar.gz}} by the \cite{Planck_2015_data_overview}, we create a covariance matrix containing the parameters we are interested in. By taking the square root of the diagonal elements corresponding to $\Omega_b$ and $h$, we get the uncertainties $\sigma_{\Omega_b} = 0.00066$ and $\sigma_{h} = 0.0057$. To add these priors to our Fisher matrices, we add $\sigma_{h}^{-2}$ and $\sigma_{\Omega_b}^{-2}$ to the corresponding diagonal values in the original matrices. We therefore neglect, for simplicity, the correlation between $h$ and $\Omega_b$ in the Planck prior. We note that our goal is not primarily to perform a realistic forecast, but to study the impact of prior choices on the numerical stability of the Fisher analysis in a simple framework.

While adding this prior reduces the largest entry-wise relative differences between the covariance matrices (for the $h-w_0$ and $n_s-w_0$ elements) from a factor of $\sim 15$  to $\sim 0.5$, it also increases other elements like the $h-10^9A_s$ element from a factor of $\sim 0.5$ to $\sim 2$. As such, we do not find that adding priors solves the problem of high entry-wise relative differences upon inversion of Fisher matrices.

Lastly, it is worth noting that the high condition numbers appear to stem from the size of our parameter space. By only varying a subset of our original parameters, we can reduce the size of the parameter space and generate new Fisher matrices using CCL and CosmoSIS. After trying various combinations of parameters, we found that once the parameter space included at least four parameters, then entry-wise relative differences between the inverted Fisher matrices started to exceed 10\%, despite much smaller entry-wise relative differences, of similar magnitude to those in Table~\ref{tab:cov_rel_diff}, in the original Fisher matrices. This agrees with idea that increasing the dimensionality of a matrix to be inverted results in this inversion being more sensitive to small deviations in the original matrix, as discussed in a Fisher forecasting in e.g. \cite{Euclidforecasts}.


\begin{deluxetable}{c|ccccc}[ht]
\tablecaption{Summary table for CCL and CosmoSIS forecasts. 
\label{tab:summary}}
\tablehead{
\colhead{Forecast}  & \colhead{Condition Number} & \colhead{$w_0-w_a$ FoM} & \colhead{$\Omega_c-10^9 A_s$ FoM} & \colhead{$\Omega_c h^2-10^9 A_s$ FoM} & \colhead{Inverse Norm Spread}
}
\startdata
CCL             & $1.07\times 10^{7}$  & $25.24$ & $96.07$ & N/A & $1.70\times 10^{-22}$ \\
CosmoSIS        & $8.08\times 10^{6}$  & $29.50$ & $221.5$ & N/A & $2.28\times 10^{-23}$ \\
CCL ($\Omega_ch^2$ \& $\Omega_bh^2$)      & $3.33\times 10^{7}$  & $26.84$ & N/A &  $70.89$ & $1.55\times 10^{-21}$ \\
\enddata
\end{deluxetable}


\begin{deluxetable}{c|ccccccc}[ht]
\tablecaption{Magnitude of relative differences for CCL and CosmoSIS Fisher matrices. The largest differences are highlighted in bold. \label{tab:fisher_rel_diff}}
\tablehead{
\colhead{} & \colhead{$\Omega_c$} & \colhead{$w_0$} & \colhead{$h$} & \colhead{$10^9 A_s$} & \colhead{$\Omega_b$} &  \colhead{$n_s$} & \colhead{$w_a$}
}
\startdata
$\Omega_c$ & 4.68E-5 & 1.92E-3 & 3.26E-4 & 5.98E-4 & 1.49E-3 & 1.01E-4 & 1.93E-4 \\
$w_0$ & 1.92E-3 & \textbf{3.84E-3} & 2.39E-3 & 1.53E-3 & 5.61E-4 & 1.36E-3 & 2.34E-3 \\ 
$h$ & 3.26E-4 & 2.39E-3 & 7.20E-4 & 1.77E-4 & 1.41E-3 & 4.82E-5 & 5.84E-4 \\
$10^9 A_s$ & 5.98E-4 & 1.53E-3 & 1.77E-4 & 1.05E-3 & 2.47E-3 & 9.57E-4 & 4.11E-4 \\
$\Omega_b$ & 1.49E-3 & 5.61E-4 & 1.41E-3 & 2.47E-3 & \textbf{5.90E-3} & 1.84E-4 & 2.51E-3 \\
$n_s$ & 1.01E-4 & 1.36E-3 & 4.82E-5 & 9.57E-4 & 1.84E-4 & 6.15E-4 & 6.83E-4 \\
$w_a$ & 1.93E-4 & 2.34E-3 & 5.84E-4 & 4.11E-4 & 2.51E-3 & 6.83E-4 & 3.32E-5 \\
\enddata
\end{deluxetable}


\begin{deluxetable}{c|ccccccc}[ht]
\tablecaption{Magnitude of relative differences for CCL and CosmoSIS covariance matrices. The largest differences are highlighted in bold. \label{tab:cov_rel_diff}}
\tablehead{
\colhead{} & \colhead{$\Omega_c$} & \colhead{$w_0$} & \colhead{$h$} & \colhead{$10^9 A_s$} & \colhead{$\Omega_b$} &  \colhead{$n_s$} & \colhead{$w_a$}
}
\startdata
$\Omega_c$ & 7.91E-1 & 1.14E+0 & 7.71E-1 & 7.36E-1 & 7.82E-1 & 6.20E-1 & 1.00E+0 \\
$w_0$ & 1.14E+0 & 3.54E-1 & \textbf{1.15E+1} & 2.88E-2 & 7.35E-1 & \textbf{1.48E+1} & 3.60E-1 \\
$h$ & 7.71E-1 & \textbf{1.15E+1} & 6.73E-1 & 4.71E-1 & 6.82E-1 & 4.20E-1 & \textbf{1.31E+1} \\
$10^9 A_s$ & 7.36E-1 & 2.88E-2 & 4.71E-1 & 2.00E-1 & 4.66E-1 & 3.17E-1 & 3.34E-2 \\
$\Omega_b$ &  7.82E-1 & 7.35E-1 & 6.82E-1 & 4.66E-1 & 6.51E-1 & 5.95E-1 & 5.55E-1 \\
$n_s$ & 6.20E-1 & \textbf{1.48E+1} & 4.20E-1 & 3.17E-1 & 5.95E-1 & 3.15E-4 & 3.45E-1 \\
$w_a$ & 1.00E+0 & 3.60E-1 & \textbf{1.31E+1} & 3.34E-2 & 5.55E-1 & 3.45E-1 & 2.99E-1 \\
\enddata
\end{deluxetable}

\section{Conclusions} \label{sec:conc}

In this note, we investigated numerical issues that may arise when Fisher forecasting LSST-like weak lensing probes to constrain dark energy, focusing on an angular power spectrum formalism whose performance was gauged by the DETF FoM. Using two numerical libraries, CCL and CosmoSIS, we generated data-vectors and data-covariances that were very close numerically, as seen by their signal-to-noise ratios having a relative difference of $-0.04$\%. However, when trying to calculate the Fisher matrix, we found that numerical instability can arise when  tuning the step-size for the derivatives that go into Equation \ref{eq:fisher}, which consequently introduces numerical instability in the DETF FoM as seen in Figure \ref{fig:cosmosis_tuning}. 
Furthermore, even with two Fisher matrices whose entry-wise relative differences were small (around 0.6\% at most), we found that these small perturbations can be magnified greatly upon inversion (to order of magnitude differences) due to these Fisher matrices having high condition numbers --- due, at least in part, to the number of parameters chosen to investigate. Our work agrees with similar findings in \cite{Euclidforecasts} and \cite{Yahia2020}, with the former estimating that for the forecast parameter covariance matrix to be stable to 10\%, the maximum error on an element of the Fisher matrix permitted would be 0.01\%.
Ensuring all parameters have a similar order of magnitude may help reduce the condition number; however, reparameterizing $\Omega_c$ by $\Omega_c h^2$ and $\Omega_b$ and $\Omega_bh^2$, or even applying priors on certain parameters does not necessarily yield significant benefits. While the large condition numbers may not necessarily propagate into large differences for the DETF FoM, as in our case, it worth carrying out similar numerical stability tests in future work with Fisher forecasts, as it may impact other quantities whose derivation involves inverting the Fisher matrix.

\acknowledgments

We thank the CCL team for making their code available at \url{https://github.com/LSSTDESC/CCL} and the CosmoSIS team for making their code available at \url{https://bitbucket.org/joezuntz/cosmosis}. We also thank Alain Blanchard, Stefano Camera, Luke Hart, Mark Kamionkowski, Eric Linder, An\u{z}e Slosar and Joe Zuntz,   for useful discussions that improved the content of this paper.

\bibliography{rn}{}

\begin{thebibliography}{}
\expandafter\ifx\csname natexlab\endcsname\relax\def\natexlab#1{#1}\fi
\providecommand{\url}[1]{\href{#1}{#1}}
\providecommand{\dodoi}[1]{doi:~\href{http://doi.org/#1}{\nolinkurl{#1}}}
\providecommand{\doeprint}[1]{\href{http://ascl.net/#1}{\nolinkurl{http://ascl.net/#1}}}
\providecommand{\doarXiv}[1]{\href{https://arxiv.org/abs/#1}{\nolinkurl{https://arxiv.org/abs/#1}}}

\bibitem[{{Aihara} {et~al.}(2018){Aihara}, {Arimoto}, {Armstrong}, {Arnouts},
  {Bahcall}, {Bickerton}, {Bosch}, {Bundy}, {Capak}, {Chan}, {Chiba}, {Coupon},
  {Egami}, {Enoki}, {Finet}, {Fujimori}, {Fujimoto}, {Furusawa}, {Furusawa},
  {Goto}, {Goulding}, {Greco}, {Greene}, {Gunn}, {Hamana}, {Harikane},
  {Hashimoto}, {Hattori}, {Hayashi}, {Hayashi}, {He{\l}miniak}, {Higuchi},
  {Hikage}, {Ho}, {Hsieh}, {Huang}, {Huang}, {Ikeda}, {Imanishi}, {Inoue},
  {Iwasawa}, {Iwata}, {Jaelani}, {Jian}, {Kamata}, {Karoji}, {Kashikawa},
  {Katayama}, {Kawanomoto}, {Kayo}, {Koda}, {Koike}, {Kojima}, {Komiyama},
  {Konno}, {Koshida}, {Koyama}, {Kusakabe}, {Leauthaud}, {Lee}, {Lin}, {Lin},
  {Lupton}, {Mand elbaum}, {Matsuoka}, {Medezinski}, {Mineo}, {Miyama},
  {Miyatake}, {Miyazaki}, {Momose}, {More}, {More}, {Moritani}, {Moriya},
  {Morokuma}, {Mukae}, {Murata}, {Murayama}, {Nagao}, {Nakata}, {Niida},
  {Niikura}, {Nishizawa}, {Obuchi}, {Oguri}, {Oishi}, {Okabe}, {Okamoto},
  {Okura}, {Ono}, {Onodera}, {Onoue}, {Osato}, {Ouchi}, {Price}, {Pyo}, {Sako},
  {Sawicki}, {Shibuya}, {Shimasaku}, {Shimono}, {Shirasaki}, {Silverman},
  {Simet}, {Speagle}, {Spergel}, {Strauss}, {Sugahara}, {Sugiyama}, {Suto},
  {Suyu}, {Suzuki}, {Tait}, {Takada}, {Takata}, {Tamura}, {Tanaka}, {Tanaka},
  {Tanaka}, {Tanaka}, {Terai}, {Terashima}, {Toba}, {Tominaga}, {Toshikawa},
  {Turner}, {Uchida}, {Uchiyama}, {Umetsu}, {Uraguchi}, {Urata}, {Usuda},
  {Utsumi}, {Wang}, {Wang}, {Wong}, {Yabe}, {Yamada}, {Yamanoi}, {Yasuda},
  {Yeh}, {Yonehara}, \& {Yuma}}]{2018PASJ...70S...4A}
{Aihara}, H., {Arimoto}, N., {Armstrong}, R., {et~al.} 2018, \pasj, 70, S4,
  \dodoi{10.1093/pasj/psx066}

\bibitem[{Albrecht {et~al.}(2006)Albrecht, Bernstein, Cahn, Freedman, Hewitt,
  Hu, Huth, Kamionkowski, Kolb, Knox, Mather, Staggs, \&
  Suntzeff}]{albrecht2006report}
Albrecht, A., Bernstein, G., Cahn, R., {et~al.} 2006, Report of the Dark Energy
  Task Force.
\newblock \doarXiv{astro-ph/0609591}

\bibitem[{Ballardini {et~al.}(2018)Ballardini, Finelli, Maartens, \&
  Moscardini}]{Ballardini_2018}
Ballardini, M., Finelli, F., Maartens, R., \& Moscardini, L. 2018, Journal of
  Cosmology and Astroparticle Physics, 2018, 044,
  \dodoi{10.1088/1475-7516/2018/04/044}

\bibitem[{{Bartelmann} \& {Schneider}(2001)}]{2001PhR...340..291B}
{Bartelmann}, M., \& {Schneider}, P. 2001, \physrep, 340, 291,
  \dodoi{10.1016/S0370-1573(00)00082-X}

\bibitem[{Bassett {et~al.}(2011)Bassett, Fantaye, Hlozek, \&
  Kotze}]{BASSETT_2011}
Bassett, B.~A., Fantaye, Y., Hlozek, R., \& Kotze, J. 2011, International
  Journal of Modern Physics D, 20, 2559–2598,
  \dodoi{10.1142/s0218271811020548}

\bibitem[{Blanchard {et~al.}(2020)Blanchard, Camera, Carbone, Cardone, Casas,
  Clesse, Ili{\'c}, Kilbinger, Kitching, Kunz, {et~al.}}]{Euclidforecasts}
Blanchard, A., Camera, S., Carbone, C., {et~al.} 2020, Astronomy \&
  Astrophysics, 642, A191

\bibitem[{{Blandford} {et~al.}(1991){Blandford}, {Saust}, {Brainerd}, \&
  {Villumsen}}]{1991MNRAS.251..600B}
{Blandford}, R.~D., {Saust}, A.~B., {Brainerd}, T.~G., \& {Villumsen}, J.~V.
  1991, \mnras, 251, 600, \dodoi{10.1093/mnras/251.4.600}

\bibitem[{Bunch(1971)}]{equilibration_sym}
Bunch, J.~R. 1971, J. ACM, 18, 566–572, \dodoi{10.1145/321662.321670}

\bibitem[{{Camera} {et~al.}(2017){Camera}, {Harrison}, {Bonaldi}, \&
  {Brown}}]{2017MNRAS.464.4747C}
{Camera}, S., {Harrison}, I., {Bonaldi}, A., \& {Brown}, M.~L. 2017, \mnras,
  464, 4747, \dodoi{10.1093/mnras/stw2688}

\bibitem[{{Chang} {et~al.}(2013){Chang}, {Jarvis}, {Jain}, {Kahn}, {Kirkby},
  {Connolly}, {Krughoff}, {Peng}, \& {Peterson}}]{eff_num_dens}
{Chang}, C., {Jarvis}, M., {Jain}, B., {et~al.} 2013, \mnras, 434, 2121,
  \dodoi{10.1093/mnras/stt1156}

\bibitem[{Chisari {et~al.}(2019)Chisari, Alonso, Krause, Leonard, Bull, Neveu,
  Villarreal, Singh, McClintock, Ellison, \& et~al.}]{CCL_2019}
Chisari, N.~E., Alonso, D., Krause, E., {et~al.} 2019, The Astrophysical
  Journal Supplement Series, 242, 2, \dodoi{10.3847/1538-4365/ab1658}

\bibitem[{Croz \& Higham(1992)}]{stab_mat_inv}
Croz, J. J.~D., \& Higham, N.~J. 1992, IMA Journal of Numerical Analysis, 12,
  1, \dodoi{10.1093/imanum/12.1.1}

\bibitem[{{Cutler} \& {Flanagan}(1994)}]{1994PhRvD..49.2658C}
{Cutler}, C., \& {Flanagan}, {\'E}.~E. 1994, \prd, 49, 2658,
  \dodoi{10.1103/PhysRevD.49.2658}

\bibitem[{{de Jong} {et~al.}(2013){de Jong}, {Verdoes Kleijn}, {Kuijken}, \&
  {Valentijn}}]{2013ExA....35...25D}
{de Jong}, J. T.~A., {Verdoes Kleijn}, G.~A., {Kuijken}, K.~H., \& {Valentijn},
  E.~A. 2013, Experimental Astronomy, 35, 25, \dodoi{10.1007/s10686-012-9306-1}

\bibitem[{Fornberg(1988)}]{fornberg1988generation}
Fornberg, B. 1988, Mathematics of computation, 51, 699

\bibitem[{{Hart} \& {Chluba}(2020)}]{2020MNRAS.495.4210H}
{Hart}, L., \& {Chluba}, J. 2020, \mnras, 495, 4210,
  \dodoi{10.1093/mnras/staa1426}

\bibitem[{Heavens {et~al.}(2014)Heavens, Seikel, Nord, Aich, Bouffanais,
  Bassett, \& Hobson}]{Heavens_2014}
Heavens, A.~F., Seikel, M., Nord, B.~D., {et~al.} 2014, Monthly Notices of the
  Royal Astronomical Society, 445, 1687–1693, \dodoi{10.1093/mnras/stu1866}

\bibitem[{{Ivezi{\'c}} {et~al.}(2019){Ivezi{\'c}}, {Kahn}, {Tyson}, {Abel},
  {Acosta}, {Allsman}, {Alonso}, {AlSayyad}, {Anderson}, {Andrew}, \&
  et~al.}]{lsst_desc_outcomes}
{Ivezi{\'c}}, {\v Z}., {Kahn}, S.~M., {Tyson}, J.~A., {et~al.} 2019, \apj, 873,
  111, \dodoi{10.3847/1538-4357/ab042c}

\bibitem[{{Jungman} {et~al.}(1996{\natexlab{a}}){Jungman}, {Kamionkowski},
  {Kosowsky}, \& {Spergel}}]{1996PhRvD..54.1332J}
{Jungman}, G., {Kamionkowski}, M., {Kosowsky}, A., \& {Spergel}, D.~N.
  1996{\natexlab{a}}, \prd, 54, 1332, \dodoi{10.1103/PhysRevD.54.1332}

\bibitem[{{Jungman} {et~al.}(1996{\natexlab{b}}){Jungman}, {Kamionkowski},
  {Kosowsky}, \& {Spergel}}]{1996PhRvL..76.1007J}
---. 1996{\natexlab{b}}, \prl, 76, 1007, \dodoi{10.1103/PhysRevLett.76.1007}

\bibitem[{{Kaiser}(1992)}]{1992ApJ...388..272K}
{Kaiser}, N. 1992, \apj, 388, 272, \dodoi{10.1086/171151}

\bibitem[{{Kirk} {et~al.}(2015){Kirk}, {Lahav}, {Bridle}, {Jouvel}, {Abdalla},
  \& {Frieman}}]{cov_donnacha_kirk}
{Kirk}, D., {Lahav}, O., {Bridle}, S., {et~al.} 2015, \mnras, 451, 4424,
  \dodoi{10.1093/mnras/stv1268}

\bibitem[{{Krause} {et~al.}(2017){Krause}, {Eifler}, {Zuntz}, {Friedrich},
  {Troxel}, {Dodelson}, {Blazek}, {Secco}, {MacCrann}, {Baxter}, {Chang},
  {Chen}, {Crocce}, {DeRose}, {Ferte}, {Kokron}, {Lacasa}, {Miranda}, {Omori},
  {Porredon}, {Rosenfeld}, {Samuroff}, {Wang}, {Wechsler}, {Abbott}, {Abdalla},
  {Allam}, {Annis}, {Bechtol}, {Benoit-Levy}, {Bernstein}, {Brooks}, {Burke},
  {Capozzi}, {Carrasco Kind}, {Carretero}, {D'Andrea}, {da Costa}, {Davis},
  {DePoy}, {Desai}, {Diehl}, {Dietrich}, {Evrard}, {Flaugher}, {Fosalba},
  {Frieman}, {Garcia-Bellido}, {Gaztanaga}, {Giannantonio}, {Gruen}, {Gruendl},
  {Gschwend}, {Gutierrez}, {Honscheid}, {James}, {Jeltema}, {Kuehn},
  {Kuhlmann}, {Lahav}, {Lima}, {Maia}, {March}, {Marshall}, {Martini},
  {Menanteau}, {Miquel}, {Nichol}, {Plazas}, {Romer}, {Rykoff}, {Sanchez},
  {Scarpine}, {Schindler}, {Schubnell}, {Sevilla-Noarbe}, {Smith},
  {Soares-Santos}, {Sobreira}, {Suchyta}, {Swanson}, {Tarle}, {Tucker},
  {Vikram}, {Walker}, \& {Weller}}]{2017arXiv170609359K}
{Krause}, E., {Eifler}, T.~F., {Zuntz}, J., {et~al.} 2017, arXiv e-prints,
  arXiv:1706.09359.
\newblock \doarXiv{1706.09359}

\bibitem[{Köhlinger {et~al.}(2017)Köhlinger, Viola, Joachimi, Hoekstra, van
  Uitert, Hildebrandt, Choi, Erben, Heymans, Joudaki, Klaes, Kuijken, Merten,
  Miller, Schneider, \& Valentijn}]{kids}
Köhlinger, F., Viola, M., Joachimi, B., {et~al.} 2017, Monthly Notices of the
  Royal Astronomical Society, 471, 4412, \dodoi{10.1093/mnras/stx1820}

\bibitem[{Laureijs {et~al.}(2011)Laureijs, Amiaux, Arduini, Augueres,
  Brinchmann, Cole, Cropper, Dabin, Duvet, Ealet, {et~al.}}]{Euclid}
Laureijs, R., Amiaux, J., Arduini, S., {et~al.} 2011, arXiv preprint
  arXiv:1110.3193

\bibitem[{{LSST Science Collaboration} {et~al.}(2009){LSST Science
  Collaboration}, {Abell}, {Allison}, {Anderson}, {Andrew}, {Angel}, {Armus},
  {Arnett}, {Asztalos}, {Axelrod}, {Bailey}, {Ballantyne}, {Bankert},
  {Barkhouse}, {Barr}, {Barrientos}, {Barth}, {Bartlett}, {Becker}, {Becla},
  {Beers}, {Bernstein}, {Biswas}, {Blanton}, {Bloom}, {Bochanski}, {Boeshaar},
  {Borne}, {Bradac}, {Brandt}, {Bridge}, {Brown}, {Brunner}, {Bullock},
  {Burgasser}, {Burge}, {Burke}, {Cargile}, {Chand rasekharan}, {Chartas},
  {Chesley}, {Chu}, {Cinabro}, {Claire}, {Claver}, {Clowe}, {Connolly}, {Cook},
  {Cooke}, {Cooray}, {Covey}, {Culliton}, {de Jong}, {de Vries}, {Debattista},
  {Delgado}, {Dell'Antonio}, {Dhital}, {Di Stefano}, {Dickinson}, {Dilday},
  {Djorgovski}, {Dobler}, {Donalek}, {Dubois-Felsmann}, {Durech},
  {Eliasdottir}, {Eracleous}, {Eyer}, {Falco}, {Fan}, {Fassnacht}, {Ferguson},
  {Fernandez}, {Fields}, {Finkbeiner}, {Figueroa}, {Fox}, {Francke}, {Frank},
  {Frieman}, {Fromenteau}, {Furqan}, {Galaz}, {Gal-Yam}, {Garnavich},
  {Gawiser}, {Geary}, {Gee}, {Gibson}, {Gilmore}, {Grace}, {Green}, {Gressler},
  {Grillmair}, {Habib}, {Haggerty}, {Hamuy}, {Harris}, {Hawley}, {Heavens},
  {Hebb}, {Henry}, {Hileman}, {Hilton}, {Hoadley}, {Holberg}, {Holman},
  {Howell}, {Infante}, {Ivezic}, {Jacoby}, {Jain}, {R}, {Jedicke}, {Jee},
  {Garrett Jernigan}, {Jha}, {Johnston}, {Jones}, {Juric}, {Kaasalainen},
  {Styliani}, {Kafka}, {Kahn}, {Kaib}, {Kalirai}, {Kantor}, {Kasliwal},
  {Keeton}, {Kessler}, {Knezevic}, {Kowalski}, {Krabbendam}, {Krughoff},
  {Kulkarni}, {Kuhlman}, {Lacy}, {Lepine}, {Liang}, {Lien}, {Lira}, {Long},
  {Lorenz}, {Lotz}, {Lupton}, {Lutz}, {Macri}, {Mahabal}, {Mandelbaum},
  {Marshall}, {May}, {McGehee}, {Meadows}, {Meert}, {Milani}, {Miller},
  {Miller}, {Mills}, {Minniti}, {Monet}, {Mukadam}, {Nakar}, {Neill}, {Newman},
  {Nikolaev}, {Nordby}, {O'Connor}, {Oguri}, {Oliver}, {Olivier}, {Olsen},
  {Olsen}, {Olszewski}, {Oluseyi}, {Padilla}, {Parker}, {Pepper}, {Peterson},
  {Petry}, {Pinto}, {Pizagno}, {Popescu}, {Prsa}, {Radcka}, {Raddick},
  {Rasmussen}, {Rau}, {Rho}, {Rhoads}, {Richards}, {Ridgway}, {Robertson},
  {Roskar}, {Saha}, {Sarajedini}, {Scannapieco}, {Schalk}, {Schindler},
  {Schmidt}, {Schmidt}, {Schneider}, {Schumacher}, {Scranton}, {Sebag},
  {Seppala}, {Shemmer}, {Simon}, {Sivertz}, {Smith}, {Allyn Smith}, {Smith},
  {Spitz}, {Stanford}, {Stassun}, {Strader}, {Strauss}, {Stubbs}, {Sweeney},
  {Szalay}, {Szkody}, {Takada}, {Thorman}, {Trilling}, {Trimble}, {Tyson}, {Van
  Berg}, {Vand en Berk}, {VanderPlas}, {Verde}, {Vrsnak}, {Walkowicz}, {Wand
  elt}, {Wang}, {Wang}, {Warner}, {Wechsler}, {West}, {Wiecha}, {Williams},
  {Willman}, {Wittman}, {Wolff}, {Wood-Vasey}, {Wozniak}, {Young}, {Zentner},
  \& {Zhan}}]{lsst_science_book}
{LSST Science Collaboration}, {Abell}, P.~A., {Allison}, J., {et~al.} 2009,
  arXiv e-prints, arXiv:0912.0201.
\newblock \doarXiv{0912.0201}

\bibitem[{Newman {et~al.}(1974)}]{newman1974determine}
Newman, M., {et~al.} 1974, J. Res. Nat. Bur. Stand., 78, 65

\bibitem[{Oliphant(2006)}]{numpy_oliphant2006}
Oliphant, T.~E. 2006, A guide to NumPy, Vol.~1 (Trelgol Publishing USA)

\bibitem[{{Planck Collaboration} {et~al.}(2016){Planck Collaboration}, {Adam,
  R.}, {Ade, P. A. R.}, {Aghanim, N.}, {Akrami, Y.}, {Alves, M. I. R.},
  {Arg\"ueso, F.}, {Arnaud, M.}, {Arroja, F.}, {Ashdown, M.}, {Aumont, J.},
  {Baccigalupi, C.}, {Ballardini, M.}, {Banday, A. J.}, {Barreiro, R. B.},
  {Bartlett, J. G.}, {Bartolo, N.}, {Basak, S.}, {Battaglia, P.}, {Battaner,
  E.}, {Battye, R.}, {Benabed, K.}, {Beno\^{\i}t, A.}, {Benoit-L\'evy, A.},
  {Bernard, J.-P.}, {Bersanelli, M.}, {Bertincourt, B.}, {Bielewicz, P.},
  {Bikmaev, I.}, {Bock, J. J.}, {B\"ohringer, H.}, {Bonaldi, A.}, {Bonavera,
  L.}, {Bond, J. R.}, {Borrill, J.}, {Bouchet, F. R.}, {Boulanger, F.},
  {Bucher, M.}, {Burenin, R.}, {Burigana, C.}, {Butler, R. C.}, {Calabrese,
  E.}, {Cardoso, J.-F.}, {Carvalho, P.}, {Casaponsa, B.}, {Castex, G.},
  {Catalano, A.}, {Challinor, A.}, {Chamballu, A.}, {Chary, R.-R.}, {Chiang, H.
  C.}, {Chluba, J.}, {Chon, G.}, {Christensen, P. R.}, {Church, S.}, {Clemens,
  M.}, {Clements, D. L.}, {Colombi, S.}, {Colombo, L. P. L.}, {Combet, C.},
  {Comis, B.}, {Contreras, D.}, {Couchot, F.}, {Coulais, A.}, {Crill, B. P.},
  {Cruz, M.}, {Curto, A.}, {Cuttaia, F.}, {Danese, L.}, {Davies, R. D.},
  {Davis, R. J.}, {de Bernardis, P.}, {de Rosa, A.}, {de Zotti, G.},
  {Delabrouille, J.}, {Delouis, J.-M.}, {D\'esert, F.-X.}, {Di Valentino, E.},
  {Dickinson, C.}, {Diego, J. M.}, {Dolag, K.}, {Dole, H.}, {Donzelli, S.},
  {Dor\'e, O.}, {Douspis, M.}, {Ducout, A.}, {Dunkley, J.}, {Dupac, X.},
  {Efstathiou, G.}, {Eisenhardt, P. R. M.}, {Elsner, F.}, {En\ss{}lin, T. A.},
  {Eriksen, H. K.}, {Falgarone, E.}, {Fantaye, Y.}, {Farhang, M.}, {Feeney,
  S.}, {Fergusson, J.}, {Fernandez-Cobos, R.}, {Feroz, F.}, {Finelli, F.},
  {Florido, E.}, {Forni, O.}, {Frailis, M.}, {Fraisse, A. A.}, {Franceschet,
  C.}, {Franceschi, E.}, {Frejsel, A.}, {Frolov, A.}, {Galeotta, S.}, {Galli,
  S.}, {Ganga, K.}, {Gauthier, C.}, {G\'enova-Santos, R. T.}, {Gerbino, M.},
  {Ghosh, T.}, {Giard, M.}, {Giraud-H\'eraud, Y.}, {Giusarma, E.}, {Gjerl\o{}w,
  E.}, {Gonz\'alez-Nuevo, J.}, {G\'orski, K. M.}, {Grainge, K. J. B.},
  {Gratton, S.}, {Gregorio, A.}, {Gruppuso, A.}, {Gudmundsson, J. E.}, {Hamann,
  J.}, {Handley, W.}, {Hansen, F. K.}, {Hanson, D.}, {Harrison, D. L.},
  {Heavens, A.}, {Helou, G.}, {Henrot-Versill\'e, S.}, {Hern\'andez-Monteagudo,
  C.}, {Herranz, D.}, {Hildebrandt, S. R.}, {Hivon, E.}, {Hobson, M.}, {Holmes,
  W. A.}, {Hornstrup, A.}, {Hovest, W.}, {Huang, Z.}, {Huffenberger, K. M.},
  {Hurier, G.}, {Ili\'{}c, S.}, {Jaffe, A. H.}, {Jaffe, T. R.}, {Jin, T.},
  {Jones, W. C.}, {Juvela, M.}, {Karakci, A.}, {Keih\"anen, E.}, {Keskitalo,
  R.}, {Khamitov, I.}, {Kiiveri, K.}, {Kim, J.}, {Kisner, T. S.}, {Kneissl,
  R.}, {Knoche, J.}, {Knox, L.}, {Krachmalnicoff, N.}, {Kunz, M.},
  {Kurki-Suonio, H.}, {Lacasa, F.}, {Lagache, G.}, {L\"ahteenm\"aki, A.},
  {Lamarre, J.-M.}, {Langer, M.}, {Lasenby, A.}, {Lattanzi, M.}, {Lawrence, C.
  R.}, {Le Jeune, M.}, {Leahy, J. P.}, {Lellouch, E.}, {Leonardi, R.},
  {Le\'on-Tavares, J.}, {Lesgourgues, J.}, {Levrier, F.}, {Lewis, A.},
  {Liguori, M.}, {Lilje, P. B.}, {Lilley, M.}, {Linden-V\o{}rnle, M.},
  {Lindholm, V.}, {Liu, H.}, {L\'opez-Caniego, M.}, {Lubin, P. M.}, {Ma,
  Y.-Z.}, {Mac\'{\i}as-P\'erez, J. F.}, {Maggio, G.}, {Maino, D.}, {Mak, D. S.
  Y.}, {Mandolesi, N.}, {Mangilli, A.}, {Marchini, A.}, {Marcos-Caballero, A.},
  {Marinucci, D.}, {Maris, M.}, {Marshall, D. J.}, {Martin, P. G.},
  {Martinelli, M.}, {Mart\'{\i}nez-Gonz\'alez, E.}, {Masi, S.}, {Matarrese,
  S.}, {Mazzotta, P.}, {McEwen, J. D.}, {McGehee, P.}, {Mei, S.}, {Meinhold, P.
  R.}, {Melchiorri, A.}, {Melin, J.-B.}, {Mendes, L.}, {Mennella, A.},
  {Migliaccio, M.}, {Mikkelsen, K.}, {Millea, M.}, {Mitra, S.},
  {Miville-Desch\^enes, M.-A.}, {Molinari, D.}, {Moneti, A.}, {Montier, L.},
  {Moreno, R.}, {Morgante, G.}, {Mortlock, D.}, {Moss, A.}, {Mottet, S.},
  {M\"unchmeyer, M.}, {Munshi, D.}, {Murphy, J. A.}, {Narimani, A.}, {Naselsky,
  P.}, {Nastasi, A.}, {Nati, F.}, {Natoli, P.}, {Negrello, M.}, {Netterfield,
  C. B.}, {N\o{}rgaard-Nielsen, H. U.}, {Noviello, F.}, {Novikov, D.},
  {Novikov, I.}, {Olamaie, M.}, {Oppermann, N.}, {Orlando, E.}, {Oxborrow, C.
  A.}, {Paci, F.}, {Pagano, L.}, {Pajot, F.}, {Paladini, R.}, {Pandolfi, S.},
  {Paoletti, D.}, {Partridge, B.}, {Pasian, F.}, {Patanchon, G.}, {Pearson, T.
  J.}, {Peel, M.}, {Peiris, H. V.}, {Pelkonen, V.-M.}, {Perdereau, O.},
  {Perotto, L.}, {Perrott, Y. C.}, {Perrotta, F.}, {Pettorino, V.},
  {Piacentini, F.}, {Piat, M.}, {Pierpaoli, E.}, {Pietrobon, D.},
  {Plaszczynski, S.}, {Pogosyan, D.}, {Pointecouteau, E.}, {Polenta, G.},
  {Popa, L.}, {Pratt, G. W.}, {Pr\'ezeau, G.}, {Prunet, S.}, {Puget, J.-L.},
  {Rachen, J. P.}, {Racine, B.}, {Reach, W. T.}, {Rebolo, R.}, {Reinecke, M.},
  {Remazeilles, M.}, {Renault, C.}, {Renzi, A.}, {Ristorcelli, I.}, {Rocha,
  G.}, {Roman, M.}, {Romelli, E.}, {Rosset, C.}, {Rossetti, M.}, {Rotti, A.},
  {Roudier, G.}, {Rouill\'e d\'{}Orfeuil, B.}, {Rowan-Robinson, M.},
  {Rubi\~no-Mart\'{\i}n, J. A.}, {Ruiz-Granados, B.}, {Rumsey, C.}, {Rusholme,
  B.}, {Said, N.}, {Salvatelli, V.}, {Salvati, L.}, {Sandri, M.}, {Sanghera, H.
  S.}, {Santos, D.}, {Saunders, R. D. E.}, {Sauv\'e, A.}, {Savelainen, M.},
  {Savini, G.}, {Schaefer, B. M.}, {Schammel, M. P.}, {Scott, D.}, {Seiffert,
  M. D.}, {Serra, P.}, {Shellard, E. P. S.}, {Shimwell, T. W.}, {Shiraishi,
  M.}, {Smith, K.}, {Souradeep, T.}, {Spencer, L. D.}, {Spinelli, M.},
  {Stanford, S. A.}, {Stern, D.}, {Stolyarov, V.}, {Stompor, R.}, {Strong, A.
  W.}, {Sudiwala, R.}, {Sunyaev, R.}, {Sutter, P.}, {Sutton, D.}, {Suur-Uski,
  A.-S.}, {Sygnet, J.-F.}, {Tauber, J. A.}, {Tavagnacco, D.}, {Terenzi, L.},
  {Texier, D.}, {Toffolatti, L.}, {Tomasi, M.}, {Tornikoski, M.}, {Tramonte,
  D.}, {Tristram, M.}, {Troja, A.}, {Trombetti, T.}, {Tucci, M.}, {Tuovinen,
  J.}, {T\"urler, M.}, {Umana, G.}, {Valenziano, L.}, {Valiviita, J.}, {Van
  Tent, F.}, {Vassallo, T.}, {Vibert, L.}, {Vidal, M.}, {Viel, M.}, {Vielva,
  P.}, {Villa, F.}, {Wade, L. A.}, {Walter, B.}, {Wandelt, B. D.}, {Watson,
  R.}, {Wehus, I. K.}, {Welikala, N.}, {Weller, J.}, {White, M.}, {White, S. D.
  M.}, {Wilkinson, A.}, {Yvon, D.}, {Zacchei, A.}, {Zibin, J. P.}, \& {Zonca,
  A.}}]{Planck_2015_data_overview}
{Planck Collaboration}, {Adam, R.}, {Ade, P. A. R.}, {et~al.} 2016, A\&A, 594,
  A1, \dodoi{10.1051/0004-6361/201527101}

\bibitem[{Raveri {et~al.}(2016)Raveri, Martinelli, Zhao, \&
  Wang}]{raveri2016information}
Raveri, M., Martinelli, M., Zhao, G., \& Wang, Y. 2016, Information Gain in
  Cosmology: From the Discovery of Expansion to Future Surveys.
\newblock \doarXiv{1606.06273}

\bibitem[{Smith {et~al.}(2003)Smith, Peacock, Jenkins, White, Frenk, Pearce,
  Thomas, Efstathiou, \& Couchman}]{Smith_2003}
Smith, R.~E., Peacock, J.~A., Jenkins, A., {et~al.} 2003, Monthly Notices of
  the Royal Astronomical Society, 341, 1311–1332,
  \dodoi{10.1046/j.1365-8711.2003.06503.x}

\bibitem[{Takahashi {et~al.}(2012)Takahashi, Sato, Nishimichi, Taruya, \&
  Oguri}]{Takahashi_2012}
Takahashi, R., Sato, M., Nishimichi, T., Taruya, A., \& Oguri, M. 2012, The
  Astrophysical Journal, 761, 152, \dodoi{10.1088/0004-637x/761/2/152}

\bibitem[{Tegmark(1997)}]{Tegmark_1997_surveys}
Tegmark, M. 1997, Physical Review Letters, 79, 3806–3809,
  \dodoi{10.1103/physrevlett.79.3806}

\bibitem[{Tegmark {et~al.}(1997)Tegmark, Taylor, \&
  Heavens}]{Tegmark_1997_fisher}
Tegmark, M., Taylor, A.~N., \& Heavens, A.~F. 1997, The Astrophysical Journal,
  480, 22–35, \dodoi{10.1086/303939}

\bibitem[{{The Dark Energy Survey Collaboration}(2005)}]{DES}
{The Dark Energy Survey Collaboration}. 2005, The Dark Energy Survey.
\newblock \doarXiv{astro-ph/0510346}

\bibitem[{{The LSST Dark Energy Science Collaboration}
  {et~al.}(2018{\natexlab{a}}){The LSST Dark Energy Science Collaboration},
  {Mandelbaum}, {Eifler}, {Hlo{\v{z}}ek}, {Collett}, {Gawiser}, {Scolnic},
  {Alonso}, {Awan}, {Biswas}, {Blazek}, {Burchat}, {Chisari}, {Dell'Antonio},
  {Digel}, {Frieman}, {Goldstein}, {Hook}, {Ivezi{\'c}}, {Kahn}, {Kamath},
  {Kirkby}, {Kitching}, {Krause}, {Leget}, {Marshall}, {Meyers}, {Miyatake},
  {Newman}, {Nichol}, {Rykoff}, {Sanchez}, {Slosar}, {Sullivan}, \&
  {Troxel}}]{lsst_srd_zenodo}
{The LSST Dark Energy Science Collaboration}, {Mandelbaum}, R., {Eifler}, T.,
  {et~al.} 2018{\natexlab{a}}, {The LSST Dark Energy Science Collaboration
  (DESC) Science Requirements Document v1 Released Data Products}, 1.0.1,
  Zenodo, \dodoi{10.5281/zenodo.2662127}

\bibitem[{{The LSST Dark Energy Science Collaboration}
  {et~al.}(2018{\natexlab{b}}){The LSST Dark Energy Science Collaboration},
  {Mandelbaum}, {Eifler}, {Hlo{\v{z}}ek}, {Collett}, {Gawiser}, {Scolnic},
  {Alonso}, {Awan}, {Biswas}, {Blazek}, {Burchat}, {Chisari}, {Dell'Antonio},
  {Digel}, {Frieman}, {Goldstein}, {Hook}, {Ivezi{\'c}}, {Kahn}, {Kamath},
  {Kirkby}, {Kitching}, {Krause}, {Leget}, {Marshall}, {Meyers}, {Miyatake},
  {Newman}, {Nichol}, {Rykoff}, {Sanchez}, {Slosar}, {Sullivan}, \&
  {Troxel}}]{lsst_srd}
---. 2018{\natexlab{b}}, arXiv e-prints, arXiv:1809.01669.
\newblock \doarXiv{1809.01669}

\bibitem[{Van Der~Walt {et~al.}(2011)Van Der~Walt, Colbert, \&
  Varoquaux}]{numpy_van2011}
Van Der~Walt, S., Colbert, S.~C., \& Varoquaux, G. 2011, Computing in Science
  \& Engineering, 13, 22

\bibitem[{Wilkinson(1961)}]{wilkinson_mat_inv}
Wilkinson, J.~H. 1961, J. ACM, 8, 281–330, \dodoi{10.1145/321075.321076}

\bibitem[{Yahia-Cherif {et~al.}(2020)Yahia-Cherif, Blanchard, Camera, Ili{\'c},
  Markovi{\v{c}}, Pourtsidou, Sakr, Sapone, \& Tutusaus}]{Yahia2020}
Yahia-Cherif, S., Blanchard, A., Camera, S., {et~al.} 2020, arXiv preprint
  arXiv:2007.01812

\bibitem[{Zuntz {et~al.}(2015)Zuntz, Paterno, Jennings, Rudd, Manzotti,
  Dodelson, Bridle, Sehrish, \& Kowalkowski}]{Zuntz_2015}
Zuntz, J., Paterno, M., Jennings, E., {et~al.} 2015, Astronomy and Computing,
  12, 45–59, \dodoi{10.1016/j.ascom.2015.05.005}

\end{thebibliography}

\end{document}